\documentclass[epj]{svjour}
\usepackage{amsmath}% amsmath package
\usepackage{graphics}
\usepackage{dcolumn}% Align table columns on decimal point
\usepackage{bm}% bold math

\newcommand{\lw}[1]{\smash{\lower 1.5ex\hbox{#1}}}
\begin{document}
\title{Analyses of $\bm{k_t}$ distributions at RHIC by means of some
selected statistical and stochastic models}
%\subtitle{Do you have a subtitle?\\ If so, write it here}
\author{M.~Biyajima\inst{1}\thanks{\emph{e-mail:}
biyajima@azusa.shinshu-u.ac.jp}
\and
M.~Kaneyama\inst{1}\thanks{\emph{e-mail:}
kaneyama@azusa.shinshu-u.ac.jp}
\and
T.~Mizoguchi$^2$\thanks{\emph{e-mail:}
mizoguti@toba-cmt.ac.jp}
\and
G.~Wilk$^3$\thanks{\emph{e-mail:}
wilk@fuw.edu.pl}
% \thanks is optional - remove next line if not needed
}                     % Do not remove
%
%\offprints{}          % Insert a name or remove this line
%
\institute{Department of Physics, Shinshu University Matsumoto 390-8621, Japan
\and Toba National College of Maritime Technology, Toba 517-8501, Japan
\and The Andrzej So\l tan Institute for Nuclear Studies, Ho\.za 69, 00681
Warsaw, Poland}
\date{Received: date / Revised version: date}
% The correct dates will be entered by Springer
%
\abstract{ The new data on $k_t$ distributions obtained at RHIC are analyzed by means
of selected models of statistical and stochastic origin in order to estimate their
importance in providing new information on hadronization process, in particular on
the value of the temperature at freeze-out to hadronic phase.
 % end of PACS codes
} %end of abstract
\maketitle

%
% SECTION-1
%

\section{Introduction} \label{intro}

\begin{figure*}
  \begin{center}
  \resizebox{0.94\textwidth}{!}{\includegraphics{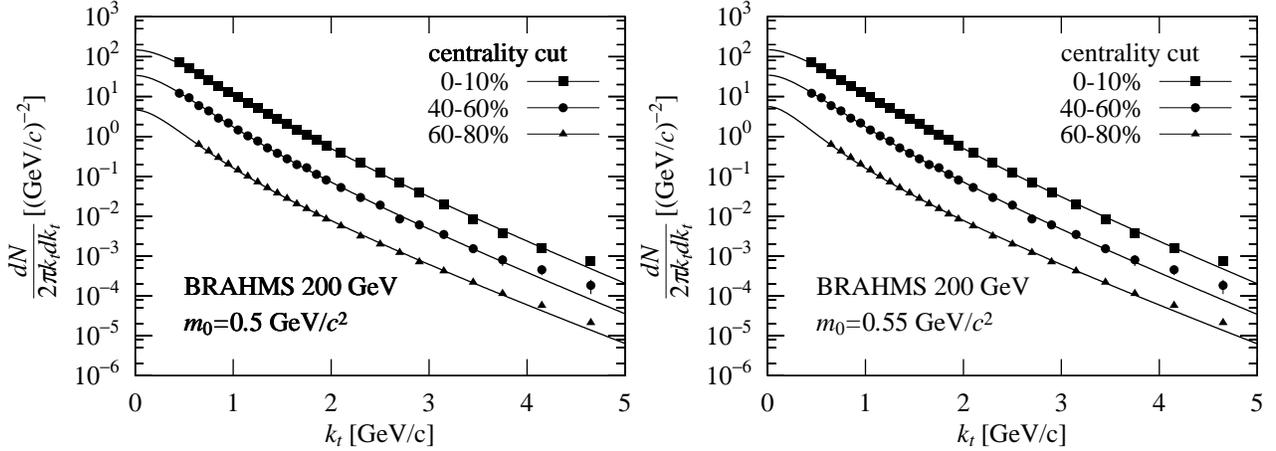}}
  \end{center}
  \caption{Results of application of simple statistical model, cf.
           Eq.~\eqref{eq:1}, to data for $k_t$-distributions at
           $\sqrt{s_{NN}} = 200$ GeV measured for different
           centralities by BRAHMS Collaboration~\cite{Arsene:2003yk}.
            }
  \label{fig:1a}
\end{figure*}

\begin{figure}
  \begin{center}
  \resizebox{0.50\textwidth}{!}{\includegraphics{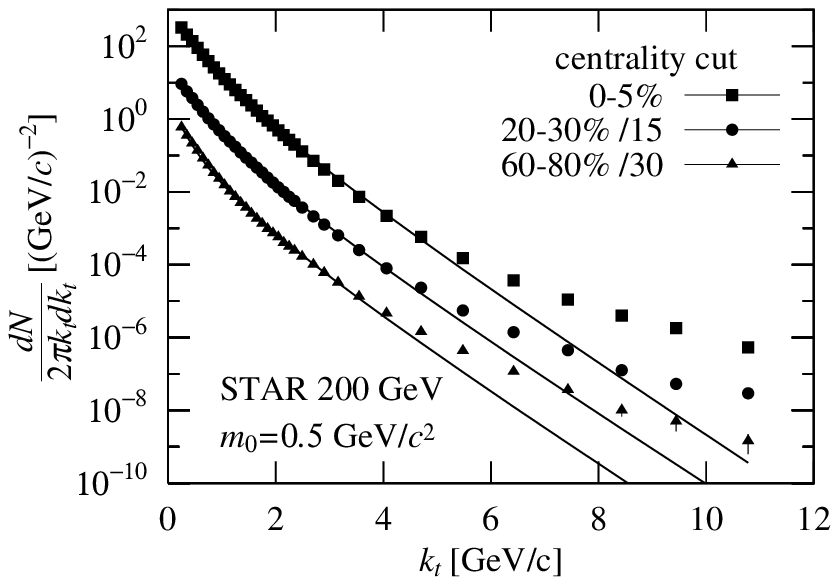}}
  \end{center}
  \caption{Results of application of simple statistical model, cf.
           Eq.~\eqref{eq:1}, to data for $k_t$-distributions at
           $\sqrt{s_{NN}} = 200$ GeV measured for different
           centralities by STAR Collaboration~\cite{Adams:2003kv}.
            }
  \label{fig:1b}
\end{figure}

Very recently high $k_t$ distributions at RHIC have been reported in
Refs.~\cite{Arsene:2003yk,Adams:2003kv,Adcox:2002pe}. These data are of potentially
high interest as a possible source of information on the conditions existing at the
freeze-out in heavy-ion collisions. This resulted in a number of works, mostly of
statistical or thermal origin \cite{explanat}, stressing different possible dynamical
aspects, like the role of resonances or the flow phenomenon. In our work we would
like to show that one can account summarily for such (and others) effects considered
in the literature by using simple minimal extensions of the known statistical or
stochastic models, which were already successfully applied in other analysis of
experimental data. They are:
\begin{itemize}
\item[$(i)$] The modified statistical model inspired by Tsallis statistics
\cite{Tsallis:1988aa}, which generalizes the usual Boltzmann-Gibbs  statistics to
nonextensive systems parametrized by a nonextensivity parameter $q$  (for
$q\rightarrow 1$ one returns to the usual Boltzmann-Gibbs extensive  scenario); it
has been already successfully used in this context \cite{Wilk,Bediaga:1999hv,Wibig}.
Parameter $q$ summarizes in such approach all deviations from the Boltzmann-Gibbs
statistics including those caused by flow phenomena and resonances \cite{explanat}.
\item[$(ii)$] A suitable adaptation of the recently proposed model derived from the
Fokker-Planck equation for the Or-stein-Uhlenbeck (O-U) process
\cite{Biyajima:2002at,B3,MC} but this time used in the transverse rapidity space,
i.e., for $y_t = \frac{1}{2}\ln [(m_t + k_t)/(m_t - k_t)]$  (where $m_t =\sqrt{m^2 +
\langle k_t\rangle^2}$), in which one allows for  the mass $m$ to be treated as free
parameter in order to account for some  specific features of data (like flow
phenomenon) which cannot be explained in a usual way.
\end{itemize}
As a kind of historical reference point we shall use classical statistical model
developed long time ago by Hagedorn \cite{Hagedorn:1965aa} in which transverse
momentum distribution of produced secondaries is given by the following formula
\cite{HR} (with $T_0$ being parameter identified with temperature, $T_h$ denoting the
so called Hagedorn temperature \cite{Hagedorn:1965aa,HR} and $m_{\pi}$ being pion
mass),
\begin{eqnarray}
\frac{d^2\sigma}{2\pi k_t dk_t} &=& C\!\! \int_{m_{\pi}}^{\infty}\!\!\!\!\! dm
\rho(m)\sqrt{m^2+k_t^2} K_1\!\!\left(\frac{\sqrt{m^2+k_t^2}}{T_0}\right)\!\!; \label{eq:1}\\
&& \rho(m) = \frac{e^{m/T_h}}{(m^2+m_0^2)^{5/4}}. \label{eq:1a}
\end{eqnarray}
As one can see in Fig. \ref{fig:1a} although fits to $k_t$ distributions at
$\sqrt{s_{NN}} = 200$ GeV obtained by BRAHMS Collaboration \cite{Arsene:2003yk} are
quite good, they start to deviate from data at highest values of $k_t$ and became
very bad there, what is very clearly seen in Fig. \ref{fig:1b} where we show our fits
to STAR data \cite{Adams:2003kv} covering larger span of transverse momenta. Although
one can argue that for such large values of $k_t$ statistical approach must give way
to some more detailed dynamical calculations \cite{explanat}, there are examples that
suitable modifications of statistical approach can lead to quite reasonable results
in leptonic, hadronic and nuclear collisions. What we have in mind here are some
non-extensive generalizations of statistical model as discussed in
\cite{Bediaga:1999hv,Wilk,Wibig}) and some specific realization of stochastic
approach as proposed by \cite{Biyajima:2002at,B3,SB}. In what follows we shall
therefore apply these two methods to nuclear data of
Refs.~\cite{Arsene:2003yk,Adams:2003kv,Adcox:2002pe}.\\

In next Section we shall analyze data using nonextensive generalization of
statistical model by means of Tsallis statistics. In Section \ref{sec:3} we shall
analyze data using stochastic approach in transverse rapidity space. Our conclusions
are presented in Section \ref{sec:4}.

\begin{table}
\caption{Values of parameters $C$, $T_h$ and $T_0$ in eq. (\ref{eq:1}) used to obtain
results presented in Figs \ref{fig:1a} and \ref{fig:1b}. The values of
$\chi^2$/n.d.f. for BRAHMS data are the same for all centralities and equal to
$19.3/23$ and $18.3/23$ for $m_0=0.5$ and $0.55$ GeV, respectively. For STAR data
they are equal $532/32$ for C.C $=0-5$\%, $249/32$ for C.C$=20-30$\% and $308/32$ for
C.C$=60-80$\%. }
\begin{center}
\begin{tabular}{|c|ccc|}
\hline
\multicolumn{4}{|c|}{BRAHMS Coll. \cite{Arsene:2003yk} $m_0=0.5$ GeV (fixed)}\\
\hline
C.C. &  $C$ & $T_h$ & $T_0$\\
(\%) &      & (GeV) & (GeV)\\
\hline
 0-10 & 177$\pm$11 &0.180$\pm$0.007
                                           & 0.169$\pm$0.006\\
10-20 & 127$\pm$9  &0.172$\pm$0.008
                                           &0.162$\pm$0.006\\
20-40 &  83$\pm$7   &0.156$\pm$0.008
                                          &0.149$\pm$0.007\\
40-60 &  44$\pm$5   &0.133$\pm$0.010
                                          &0.128$\pm$0.009\\
60-80 & 177$\pm$11 &0.095$\pm$0.0001
                                          &0.093$\pm$0.0001\\
\hline
\multicolumn{4}{|c|}{BRAHMS Coll. \cite{Arsene:2003yk} $m_0=0.55$ GeV (fixed)}\\
\hline
0-10 &  204$\pm$13  &0.172$\pm$0.008
                                           &0.162$\pm$0.006\\
10-20 & 146$\pm$10 &0.163$\pm$0.008
                                           &0.155$\pm$0.006\\
20-40 &  96$\pm$8   &0.148$\pm$0.008
                                           &0.142$\pm$0.007\\
40-60 &  51$\pm$6   &0.124$\pm$0.010
                                           &0.121$\pm$0.009\\
60-80 & 204$\pm$13 &0.075$\pm$0.00007
                                           &0.075$\pm$0.0001\\
\hline
\multicolumn{4}{|c|}{STAR Coll. \cite{Adams:2003kv} $m_0=0.5$ GeV (fixed)}\\
\hline
 0-5  &  816$\pm$15 &0.086$\pm$0.0001
                                          &0.085$\pm$0.0001\\
20-30 &  382$\pm$7  &0.077$\pm$0.0001
                                          &0.076$\pm$0.0001\\
60-80 &  106$\pm$2  &0.037$\pm$0.00001
                                          &0.037$\pm$0.00001\\
\hline
\end{tabular}
\label{tab:a2}
\end{center}
\end{table}
%
%
% SECTION-2
%

\section{Analysis of $k_t$ distributions by generalized statistical
model based on Tsallis statistics} \label{sec:2}

In many fields in physics, which use statistical and stochastic approaches as their
tools, it was recognized since some time already that the usual Boltzmann-Gibbs
approach encounters serious problems when applied to systems possessing memory
effects, correlations (especially long-range  correlations but also those caused by
the production of resonances in multiparticle production processes or by the flow
effects present there) or which phase space has some (multi) fractal structure
\cite{Tsallis:1988aa}. Such systems are all, in a sense, \textit{small}, by what we
means that the effective range of correlations they experience is of the order of
dimension of the system itself. Therefore they will not show property of extensivity
leading to Boltzmann-Gibbs form of entropy, which is the basis of any statistical or
stochastical model. One can therefore argue that in such cases one has to resort to
some dynamical approach in which effects mentioned above would be properly accounted
for. The problem is, however, that there is no unique model of this type and usually
several approaches are competing among themselves in describing experimental data.
The other possibility is to realize that most probably our system is not extensive
(in the abovementioned sense) and that this fact should be accounted for by using the
non-extensive form of entropy, for example the so called Tsallis
\cite{Tsallis:1988aa}. It turns out that such situations are encountered also in
domain of multiparticle production processes at high energy collisions (cf.,
\cite{Wilk}, to which we refer for all details). In fact, there already exists a
number of detailed analysis using a non-extensive approach ranging from $k_t$
distributions in $e^+e^-$ annihilations \cite{Bediaga:1999hv} and in $p+\bar{p}$
collisions \cite{Wibig} to rapidity distributions in some selected reactions
\cite{Wilk}. In \cite{Bediaga:1999hv,Wibig} a kind of non-extensive $q$-version of
Hagedorn approach has been used whereas \cite{Wilk} were exploring information
theoretical approach to statistical models including as option also its non-extensive
version \footnote{It should be mentioned at this point that proper formulation of
Hagedorn model using Tsallis $q$-statistics has been proposed in \cite{qHag}. We
shall not pursue this problem here.}.

In our work we shall apply Tsallis formalism, treated as simplest possible extension
of the usual statistical approach with parameter $q$ (the so called nonextensivity
parameter or entropic index) summarizing deviations from the usual statistical
approach (without, however, specifying their dynamical origin). It leads to ($T_0$
denotes temperature):
\begin{eqnarray}
\hspace{-7mm}\frac{d^2\sigma}{2\pi k_t dk_t} = C\int_0^{\infty}\left[ 1 -
(1-q)\frac{\sqrt{k_t^2+k_l^2 + m^2}}{T_0}
\right]^{Q}\, dk_l.
\label{eq:2}
\end{eqnarray}
There exist two different formulations leading to slightly different
forms of parameter $Q$:
\begin{itemize}
\item[$(a)$] In first one uses the so-called escort probability
distributions \cite{escort}, $P_i = p_i^q/\sum_i p_i^q$ (cf., for
example, analysis of $k_t$ distributions in $e^+e^-$ annihilations
\cite{Bediaga:1999hv} or in $p\bar{p}$ collisions\cite{Wibig}), in this
case $Q = q/(1-q)$.
\item[$(b)$] In second approach one uses normal definition of
probabilities resulting in $Q = 1/(1-q)$. In this case, as shown in
\cite{Wilk:1999dr,ql1}, parameter $q$ is given by the normalized
variance of all intrinsic fluctuations present in the hadronizing
system under consideration:
\begin{equation}
q = 1 + \omega =1 + \left( \langle \beta^2\rangle - \langle \beta
\rangle^2\right)/\langle \beta\rangle^2 . \label{eq:qq}
\end{equation}
This conjecture originates from the observation that:
\begin{eqnarray}
[1-(1-q)\beta_0 H_0]^{\frac {1}{1-q}} = \int_0^{\infty}e^{-\beta H_0}
f(\beta)d\beta
\label{eq:6}
\end{eqnarray}
where $f(\beta)$ describes fluctuation of parameter $\beta$  and has form of Gamma
distribution \cite{Wilk:1999dr,ql1} (in our case $H_0=\sqrt{k_l^2+k_t^2+m^2}$ and
fluctuations are in temperature, i.e.,  $\beta =1/T$ and $\beta_0 =\langle \beta
\rangle$ with respect to $f(\beta)$) \footnote{It must be mentioned at this point
that this suggestion, which in \cite{Wilk:1999dr} has been derived only for $q>1$
case, has been shown to be valid also for $q<1$ case \cite{ql1} and extended to
general form of fluctuations leading then to the new concept of
\textit{superstatistics} proposed in \cite{superq}. The most recent discussion of
physical meaning of $q$ parameter when applied to multiparticle production processes
(and in this context also of the possible origin of statistical formulas as well)
with relevant references can be found in \cite{Wilk}.}.
\end{itemize}
\begin{figure*}
  \begin{center}
  \resizebox{0.94\textwidth}{!}{\includegraphics{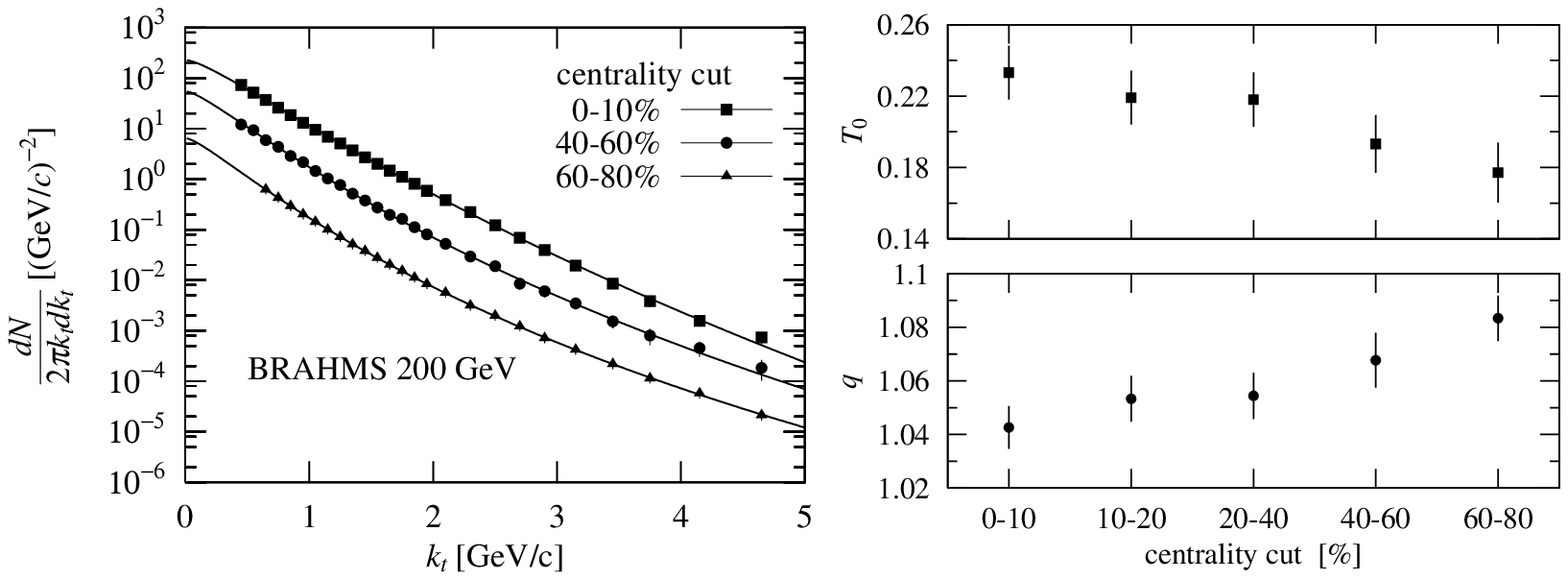}}
  \resizebox{0.94\textwidth}{!}{\includegraphics{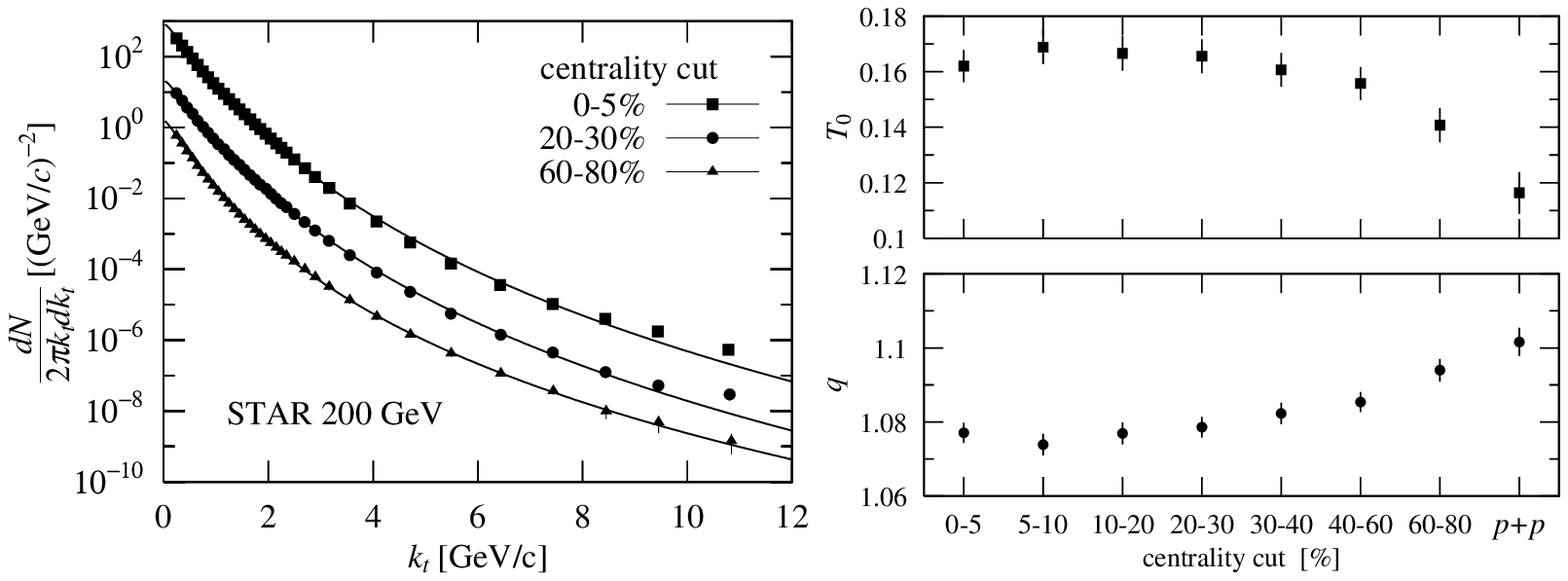}}
  \resizebox{0.94\textwidth}{!}{\includegraphics{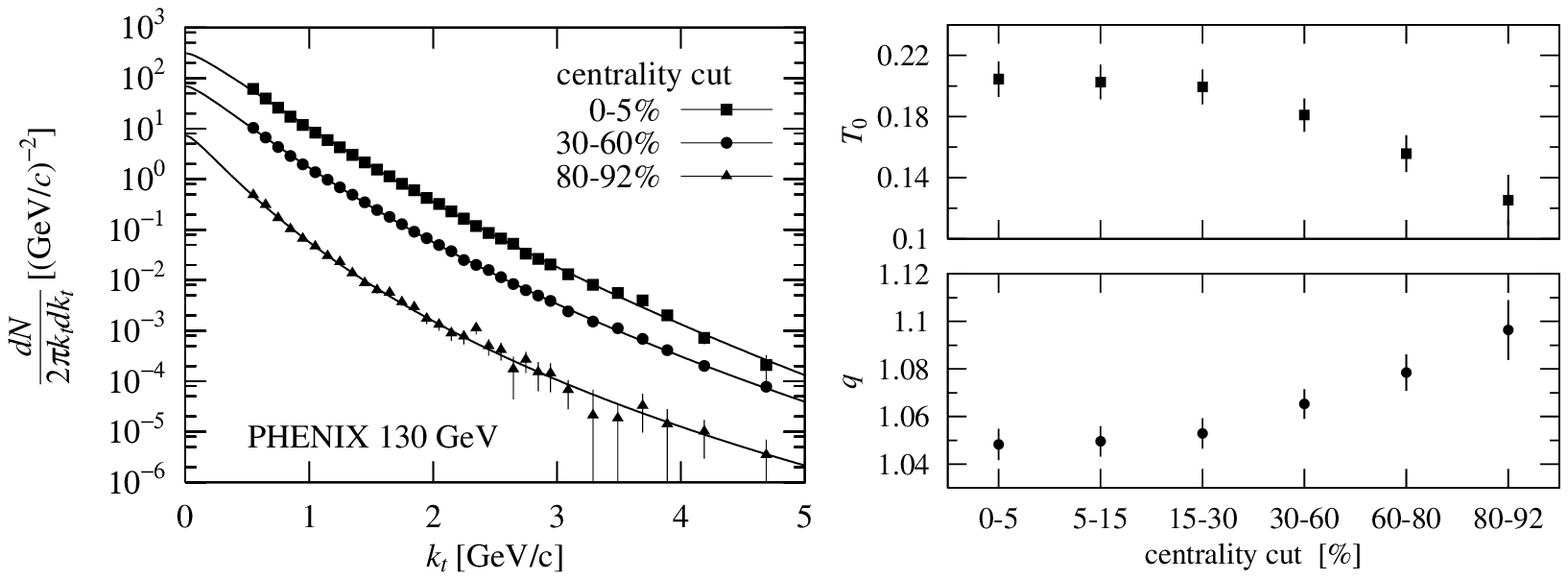}}
  \end{center}
  \caption{Results of application of non-extensive approach given by
           Eq.~\eqref{eq:2} with $Q=q/(1-q)$ to data for $k_t$-distributions
           at $\sqrt{s_{NN}} = 200$ GeV measured for different
           centralities by BRAHMS \cite{Arsene:2003yk} and STAR
           \cite{Adams:2003kv} Collaborations and to data at
           $\sqrt{s_{NN}}=130$ GeV as measured by PHENIX Collaboration
           \cite{Adcox:2002pe}.
           The results obtained using $Q=1/(1-q)$ instead look
           essentially the same, therefore they are not shown separately.
           For differences in values of obtained parameters see
           Table~\ref{table:q-results}.
           }
  \label{fig:2}
\end{figure*}

%%% TABLE 1 %%%

%
% For tables use
\begin{table*}
\caption{Values of characteristic parameters used to fit data on $k_t$-distributions
at different centralities by using non-extensive approach as given by
Eq.~\eqref{eq:2} with $Q=q/(1-q)$ and $Q=1/(1-q)$ for data at $\sqrt{s_{NN}} = 200$
GeV obtained by BRAHMS \cite{Arsene:2003yk} and STAR \cite{Adams:2003kv}
Collaborations and at $\sqrt{s_{NN}} = 130$ GeV obtained by PHENIX Collaboration
\cite{Adcox:2002pe}. For results obtained using $Q=1/(1-q)$ we provide also explicit
values of the corresponding fluctuations of temperature as given by $\Delta
T_0=\sqrt{q-1}\cdot T_0$. The order of magnitude of the corresponding errors for
$T_0$, $q$ and $\Delta T_0$: $\delta T_0$, $\delta q$ and $\delta \Delta T_0$, ,
respectively, are listed below as well. In analysis we have used errors either as
provided by experiments (for STAR and PHENIX) or assuming systematic error on the
level of $5$\% (for BRAHMS). }
\begin{center}
\begin{tabular}{|c|cccc|ccccc|}
\hline
%\multicolumn{10}{|c|}{}\\
\multicolumn{10}{|c|}{BRAHMS Coll. \cite{Arsene:2003yk}}\\
%\multicolumn{10}{|c|}{}\\
%&&&&&&&&&\\
\hline
       &\multicolumn{4}{|c|}{ Eq.~\eqref{eq:2} with $Q=q/(1-q)$}
&\multicolumn{5}{|c|}{ Eq.~\eqref{eq:2} with $Q=1/(1-q)$}\\
       &\multicolumn{4}{|c|}{$\delta T_0=$0.005-0.007, $\delta q
=$0.003-0.005.} &\multicolumn{5}{|c|}{$\delta T_0 =$0.005-0.007,
$\delta q=$0.002-0.004,}\\
       &\multicolumn{4}{|c|}{}&\multicolumn{5}{|c|}{ $\delta \Delta
T_0 =$0.001-0.002.}\\
\hline
C.C.   & $\chi^2$/n.d.f. & C & $T_0$ & $q$ & $\chi^2$/n.d.f. & C &
$T_0$ & $q$ & $\Delta T_0$ \\
(\%)   &                 &   & (GeV) &     &                 &   &
(GeV) &     & (GeV)      \\
\hline
 0-10 & 11.2/23 & 1033$\pm$78 & 0.232 & 1.043 & 11.2/23 & 1033$\pm$78 &
  0.223 & 1.041 & 0.045\\
10-20 & 12.9/23 & 797$\pm$66  & 0.224 & 1.049 & 12.9/23 & 797$\pm$65  &
 0.213 & 1.047 & 0.046\\
20-40 & 12.7/23 & 525$\pm$49  & 0.215 & 1.055 & 12.7/23 & 525$\pm$49  &
 0.204 & 1.052 & 0.047\\
40-60 & 10.5/23 & 304$\pm$38  & 0.193 & 1.067 & 10.5/23 & 304$\pm$38  &
 0.181 & 1.063 & 0.045\\
60-80 & 2.85/22 & 41$\pm$5    & 0.175 & 1.084 & 2.85/22 & 41$\pm$5    &
 0.161 & 1.077 & 0.045\\
\hline
\hline
%\multicolumn{10}{|c|}{}\\
\multicolumn{10}{|c|}{STAR Coll. \cite{Adams:2003kv}}\\
%\multicolumn{10}{|c|}{}\\
%&&&&&&&&&\\
\hline
       &\multicolumn{4}{|c|}{ Eq.~\eqref{eq:2} with $Q=q/(1-q)$}
&\multicolumn{5}{|c|}{ Eq.~\eqref{eq:2} with $Q=1/(1-q)$}\\
       &\multicolumn{4}{|c|}{ $\delta T_0 =$0.002-0.003, $\delta q=$
0.001-0.002.} &\multicolumn{5}{|c|}{$\delta T_0=$0.002-0.003, $\delta
q\cong$0.001,}\\
       &\multicolumn{4}{|c|}{} &\multicolumn{5}{|c|}{$\delta \Delta
T_0\cong$0.001.}\\
\hline
C.C.   & $\chi^2$/n.d.f. & C & $T_0$ & $q$ & $\chi^2$/n.d.f. & C &
$T_0$ & $q$ & $\Delta T_0$ \\
(\%)   &                 &   & (GeV) &     &                 &   &
(GeV) &     & (GeV)      \\
\hline
0-5   & 170/32 & 4684$\pm$231 & 0.171 & 1.071 & 170/32  & 4686$\pm$231 &
 0.159 & 1.066 & 0.041\\
5-10  &  68/32 & 3393$\pm$184 & 0.176 & 1.068 & 67.8/32 & 3393$\pm$185 &
 0.165 & 1.064 & 0.041\\
10-20 &  69/32 & 2767$\pm$144 & 0.171 & 1.073 & 69.2/32 & 2768$\pm$144 &
 0.160 & 1.068 & 0.042\\
20-30 &  45/32 & 1928$\pm$102 & 0.169 & 1.075 & 44.7/32 & 1928$\pm$102 &
 0.157 & 1.070 & 0.042\\
30-40 &  44/32 & 1391$\pm$78  & 0.165 & 1.078 & 43.9/32 & 1391$\pm$78  &
 0.153 & 1.072 & 0.041\\
40-60 &  19/32 &  896$\pm$50  & 0.153 & 1.085 & 19.2/32 &  896$\pm$50  &
 0.141 & 1.079 & 0.040\\
60-80 &  14/32 &  414$\pm$25  & 0.137 & 1.095 & 14.2/32 &  413$\pm$25  &
 0.125 & 1.087 & 0.037\\
$p+p$ & 9.7/29 &   62$\pm$7   & 0.117 & 1.099 & 9.62/29 & 61.9$\pm$7.1 &
 0.107 & 1.090 & 0.032\\
\hline
\hline
%\multicolumn{10}{|c|}{}\\
\multicolumn{10}{|c|}{PHENIX Coll. \cite{Adcox:2002pe}}\\
%\multicolumn{10}{|c|}{}\\
%&&&&&&&&&\\
\hline
       &\multicolumn{4}{|c|}{ Eq.~\eqref{eq:2} with $Q=q/(1-q)$}
&\multicolumn{5}{|c|}{ Eq.~\eqref{eq:2} with $Q=1/(1-q)$}\\
       &\multicolumn{4}{|c|}{ $\delta T_0=$0.011-0.016, $\delta
q=$0.008-0.010.} &\multicolumn{5}{|c|}{ $\delta T_0=$0.011-0.016,
$\delta q=$0.005-0.011,}\\
       &\multicolumn{4}{|c|}{} &\multicolumn{5}{|c|}{ $\delta \Delta
T_0=$0.003-0.005.}\\
\hline
C.C.   & $\chi^2$/n.d.f. & C & $T_0$ & $q$ & $\chi^2$/n.d.f. & C &
$T_0$ & $q$ & $\Delta T_0$ \\
(\%)   &                 &   & (GeV) &     &                 &   &
(GeV) &     & (GeV)      \\
\hline
0-5   & 5.13/29 & 1694$\pm$409 & 0.201 & 1.049 & 5.1/29  & 1694$\pm$411 &
 0.192 & 1.047 & 0.042\\
5-15  & 3.62/29 & 1330$\pm$316 & 0.199 & 1.051 & 3.6/29  & 1330$\pm$316 &
 0.190 & 1.048 & 0.042\\
15-30 & 5.55/29 &  846$\pm$206 & 0.196 & 1.054 & 5.6/29  &  846$\pm$206 &
 0.186 & 1.051 & 0.042\\
30-60 & 2.63/29 &  433$\pm$113 & 0.178 & 1.066 & 2.6/29  &  433$\pm$113 &
 0.167 & 1.074 & 0.045\\
60-80 & 10.6/29 &  139$\pm$48  & 0.152 & 1.080 & 10.6/29 &  139$\pm$48  &
 0.141 & 1.062 & 0.035\\
80-92 & 9.10/29 &   74$\pm$45  & 0.121 & 1.098 & 9.1/29  &   74$\pm$42  &
 0.110 & 1.089 & 0.033\\
\hline
\end{tabular}
\label{table:q-results}
\end{center}
\end{table*}

We have analyzed BRAHMS \cite{Arsene:2003yk}, STAR
\cite{Adams:2003kv} and PHE\-NIX \cite{Adcox:2002pe} data and our
results are shown in Fig.~\ref{fig:2} and in
Table~\ref{table:q-results}. It turns out that both form of parameter $Q$
in Eq.~\eqref{eq:2} result in practically identical curves, therefore here
we are showing only results obtained for $Q=q/(1-q)$. The values of
parameters are also very close to each other with tendency of $C$,
$T_0$ and $q$ estimated by using $Q=1/(1-q)$ being slightly
bigger then those obtained for $Q=q/(1-q)$. It is worth to
stress at this point that such comparison of these two approaches has
been made for the first time here and, as one can see from the
presented results, it confirms previous expectation (made in
\cite{Wilk}) that in case of only limited phenomenological
applications, as is the case of our work, the results from using
Eq.~\eqref{eq:2} with $Q=q/(1-q)$ (i.e., parameters: $C^{(a)} = c$,
$T_0^{(a)}=l$ and $q^{(a)}=q$) are simply connected
to those using $Q=1/(1-q)$ (i.e., to parameters: $C^{(b)} = C$,
$T_0^{(b)}=L$ and $q^{(b)} = \hat{Q}$), namely:
\begin{equation}
\hat{Q} \simeq 1-\frac{1-q}{q},\qquad L\simeq \frac{l}{q},\qquad C\simeq cq
. \label{eq:Qq}
\end{equation}
As one can see from Table~\ref{table:q-results} these relations are indeed satisfied
(some small differences present could be attributed to the fact that both sets of
results represent results of separate and independent fitting procedures, without
making use of Eq.~\eqref{eq:Qq}). It means therefore that in all phenomenological
applications one can use either of the two form of parameter $Q$ in
Eqs.~\eqref{eq:2}, and, if necessary, to use Eq.~\eqref{eq:Qq} to translate results
from one scheme to another. In both cases the pion mass value has been used, $m=0.14$
GeV (and we have checked that additional changes in mass $m$ of the type introduced
recently in \cite{SB}, would not affect the final results as long as $m$ is limited
to, say, $m<0.2$ GeV). The estimated fluctuations of temperature are of the order of
$30-45$ MeV. It is interesting to observe that these fluctuations are weaker at small
centralities and grow for more peripheral collisions matching very nicely similar
fluctuations seen in $p+p$ data \cite{Adams:2003kv} shown here for comparison. One
should add here also result from similar analysis of $e^+e^-$ data
\cite{Bediaga:1999hv} reporting even higher values of nonextensivity parameter $q$
(reaching value of $q\simeq 1.2$), i.e., much stronger fluctuations. These results
confirm therefore, for the first time, another expectation made in \cite{Wilk} saying
that precisely such trend should be observed. This is because Eq.~\eqref{eq:qq} can
be also interpreted as being a measure of the \textit{total heat capacity} $C_h$ of
the hadronizing system (cf. \cite{Wilk}):
\begin{equation}
\frac{1}{C_h} = \frac{\sigma^2(\beta)}{\langle \beta\rangle^2} = \omega
= q-1 . \label{eq:cap}
\end{equation}
As the heat capacity $C_h$ is proportional to the volume, $C_h \sim V$,
in our case $V$ would be the volume of interaction (or
hadronization),  it is expected to grow with volume and,
respectively, $q$ is expected to decrease with $V$, which is indeed
the case if one puts together results for $e^+e^-$, $p\bar{p}$ and
$AA$ collisions.

%
% SECTION-3
%

\section{Analysis of $\bm{k_t}$ distributions using stochastic approach in
$\bm{y_t}$ space}  \label{sec:3}

Whereas previous approach was concerned with extension of the purely statistical
approach the one presented now will go bit further by modelling hadronization process
by a kind of diffusion mechanism Refs.~\cite{Biyajima:2002at,B3,SB} in which the
original energy of projectiles is being dissipated in some well defined way into a
number of produced secondaries occurring in different part of the phase space
\footnote{It should be mentioned here that there exist also non-extensive versions of
such diffusion process applied to multiparticle production data \cite{qFP} but we
shall not pursue this possibility here.}. In the case considered here it is diffusion
process taking place in the $k_t$ space. Actually, it turns out that it is more
suitable to consider such diffusion as taking place in the $y_t = \sinh^{-1} (k_t/m)$
space. In this case one obtains the following Fokker-Planck equation:
\begin{eqnarray}
\frac{\partial P_t (y_t,t)}{\partial t}=\gamma \left[\frac{\partial
y_tP_t(y_t,\:t)}{\partial y_t}+\frac{\sigma_t^2}{2\gamma}\frac{\partial^2
P_t(y_t,t)}{\partial y_t^2}\right], \label{eq:3}
\end{eqnarray}
Its solution can be expressed by Gaussian distribution,
\begin{eqnarray}
\frac{d^2\sigma}{2\pi k_t dk_t} = CP_t (y_t,t) = \frac C{\sqrt{2\pi
V_t^2(t)}}\exp\left[-\frac{y_t^2}{2V_t^2(t)}\right] ,
\label{eq:4}
\end{eqnarray}
with\footnote{See also \cite{Biyajima:2002at}. Actually, Eq.
(\ref{eq:4}) is the same as the formula used already long time ago in
\cite{MC}.}
\begin{equation}
2V_t^2(t) = \frac{\sigma_t^2}{\gamma}(1-e^{-2\gamma t}) .
\end{equation}

\begin{figure*}
  \begin{center}
  \resizebox{0.94\textwidth}{!}{\includegraphics{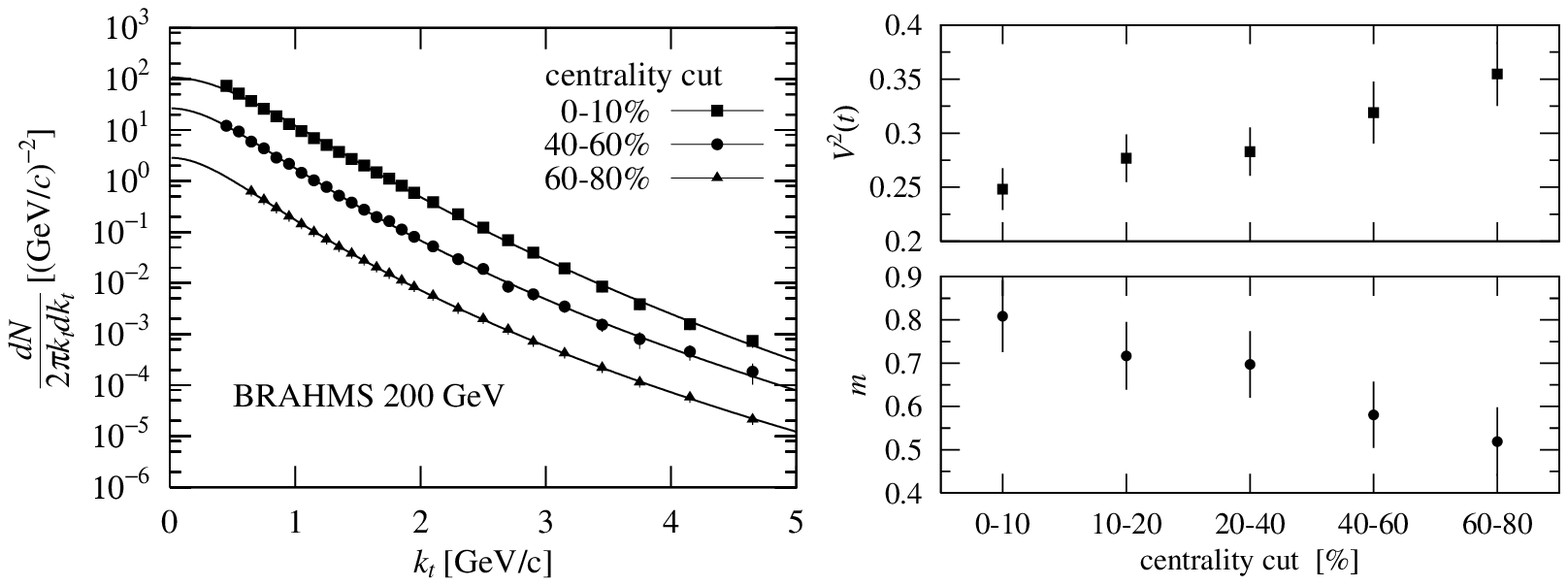}}
  \resizebox{0.94\textwidth}{!}{\includegraphics{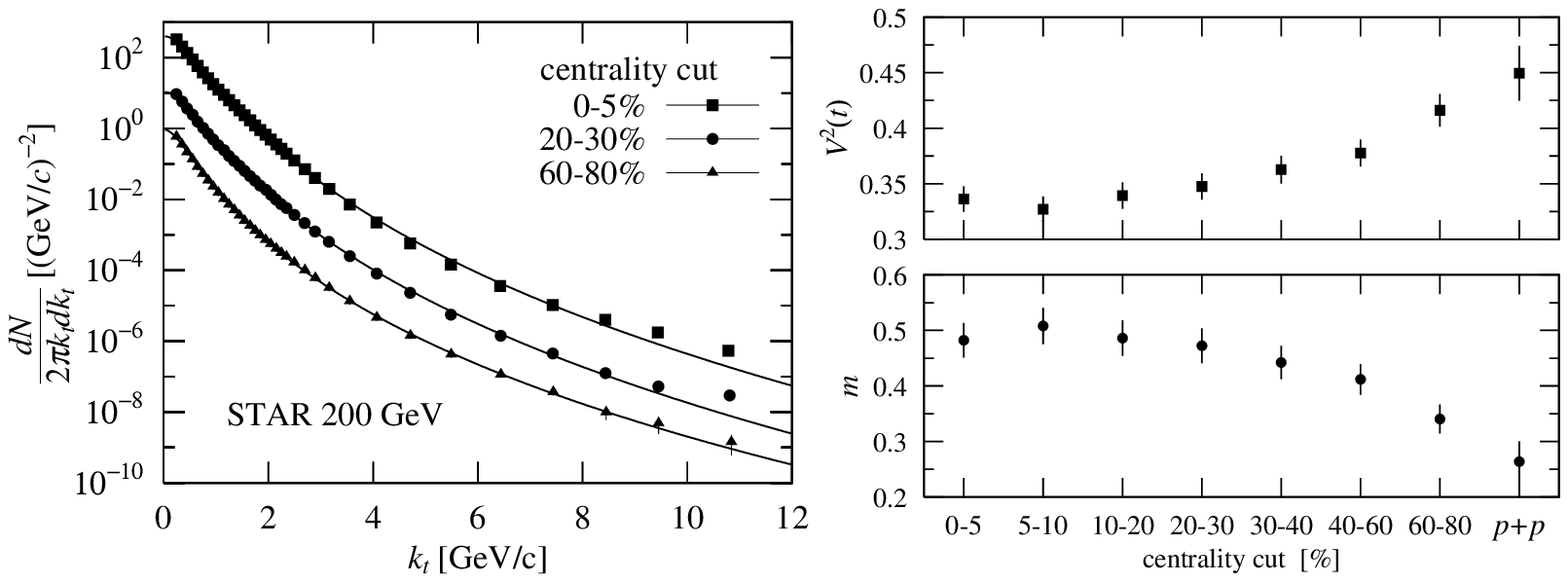}}
  \resizebox{0.94\textwidth}{!}{\includegraphics{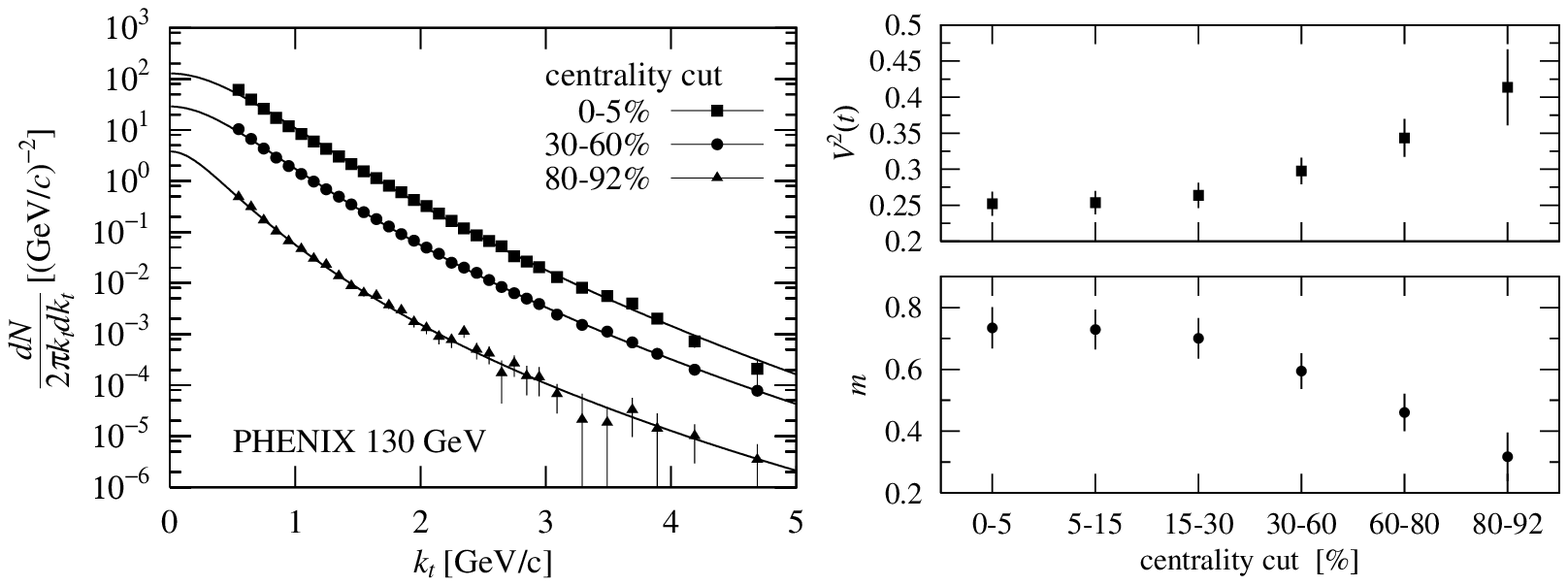}}
  \end{center}
  \caption{Results of application of stochastic approach as given by
           Eq.~\eqref{eq:4} to data for $k_t$-distributions at
           $\sqrt{s_{NN}} = 200$ GeV measured for different
           centralities by BRAHMS~\cite{Arsene:2003yk} and
           STAR~\cite{Adams:2003kv} Collaborations and at
           $\sqrt{s_{NN}} = 130$ GeV obtained by PHENIX Collaboration
           \cite{Adcox:2002pe}. Notice that mass $m$ is treated here
           as free parameter, in similar way as in \cite{SB}.}
  \label{fig:3}
\end{figure*}

%%% TABLE 2 %%%

\begin{table}
\caption{Values of characteristic parameters used to fit data on
$k_t$-distributions at different centralities by using stochastic
approach as given by Eq.~\eqref{eq:4} and presented in
Fig.~\ref{fig:3} for data at $\sqrt{s_{NN}} = 200$ GeV obtained by
BRAHMS \cite{Arsene:2003yk} and STAR \cite{Adams:2003kv}
Collaborations and at $\sqrt{s_{NN}} = 130$ GeV obtained by PHENIX
Collaboration \cite{Adcox:2002pe}. The order of magnitude of the
corresponding errors for $T_0$, $\delta T_0$, are listed below as
well.}
\begin{center}
\begin{tabular}{ccccc}
\hline
&&&\\\multicolumn{5}{c}{BRAHMS Coll.;  $\delta
T_0=$ 0.008-0.012; $\delta m =$ 0.024-0.031}\\
C.\ C.\ (\%) & $\chi^2$/n.d.f. & $C$ & $T_0$ (GeV) & $m$ (GeV) \\
\hline\
0-10  & 39.9/23 & 140$\pm$9   & 0.201 & 0.784\\
10-20 & 24.2/23 & 108$\pm$7   & 0.199 & 0.725\\
20-40 & 20.1/23 &  72$\pm$5   & 0.196 & 0.671\\
40-60 & 11.9/23 &  38$\pm$4   & 0.185 & 0.577\\
60-80 & 4.06/22 & 4.3$\pm$0.5 & 0.184 & 0.515\\
\hline
\hline
&&&\\
\multicolumn{5}{c}{STAR Coll.; $\delta T_0=$ 0.004-0.006; $\delta m
=$ 0.009-0.014} \\
C.\ C.\ (\%) & $\chi^2$/n.d.f. & $C$ & $T_0$ (GeV) & $m$ (GeV) \\
\hline
0-5   & 221/32  &   484$\pm$22   & 0.169 & 0.533\\
5-10  & 124/32  &   370$\pm$18   & 0.170 & 0.547\\
10-20 & 121/32  &   310$\pm$14   & 0.168 & 0.513\\
20-30 & 92.9/32 &   217$\pm$10   & 0.168 & 0.498\\
30-40 & 89.9/32 &   158$\pm$8    & 0.165 & 0.473\\
40-60 & 43.9/32 &  99.6$\pm$5.0  & 0.157 & 0.419\\
60-80 & 22.3/32 &  43.2$\pm$2.3  & 0.143 & 0.349\\
$p+p$ & 17.4/29 &  5.29$\pm$0.63 & 0.126 & 0.298\\
\hline
\hline
&&&\\
\multicolumn{5}{c}{PHENIX Coll.; $\delta
T_0=$ 0.020-0.037;  $\delta m =$ 0.058-0.078} \\
C.\ C.\ (\%) & $\chi^2$/n.d.f. & $C$ & $T_0$ (GeV) & $m$ (GeV) \\
\hline
0-5  & 8.06/29 & 161$\pm$34     & 0.185 & 0.734 \\
5-15 & 5.61/29 & 124$\pm$26     & 0.185 & 0.729 \\
15-30& 7.27/29 & 80$\pm$18      & 0.185 & 0.700 \\
30-60& 3.49/29 & 39$\pm$9       & 0.177 & 0.594 \\
60-80& 11.1/29 & 12$\pm$4       & 0.158 & 0.460 \\
80-92& 9.01/29 & 6.1$\pm$3.6    & 0.131 & 0.317 \\
\hline
\end{tabular}
\label{table:stoch}
\end{center}
\end{table}

In Fig.~\ref{fig:3} we show our results of using Eq.~\eqref{eq:4}. It should be
noticed that now, following \cite{SB}, we have regarded mass $m$ to be a free
parameter. Only then we can obtain good agreement with data. In a sense, variable
mass $m$ corresponds in this approach to the non-extensivity parameter $q$ introduced
in Section \ref{sec:2} in that it summarily accounts for some additional effects not
accounted by simple diffusion process (like, for example, effect of resonances and
flow).

In the stochastic approach considered here we do not have direct
access to the temperature $T_0$. It is accessible only if we
additionally assume validity of the Einstein's
fluctuation-dissipation relation, which in our case means that
measure of the size of diffusion (dissipation), $V^2_t(t)$, can be
expressed by the temperature  $T_o$ and mass $m$:
\begin{equation}
V^2(t) = \frac{T_0}{m} . \label{eq:fdt}
\end{equation}
Therefore our results for $V^2$ shown in Fig.~\ref{fig:3} (see inlets), where
$V(t)^2$ increases with increasing centrality, would indicate that temperature $T_0$
obtained by applying Einstein's relation with $m$ kept constant would increases with
centrality as well, contrary to what has been obtained above by applying
$q$-statistical approach. We have allowed then (following \cite{SB}) the  mass $m$ to
be a free parameter and the best fit is obtained when $m$ decreases with centrality,
see inlets in Fig.~\ref{fig:3}. The resulting temperature, $T_0 \simeq m\cdot V_t^2$,
behaves then in essentially the same way as function of centrality as in the
$q$-statistical approach, cf., Table~\ref{table:stoch} and
Fig.~\ref{fig:4}\footnote{It should be stressed here that for constant value of mass,
$m=0.14$ GeV as used for $q$-statistics case above, we would have obtained somewhat
higher values of $\chi^2$'s. In addition, it is interesting to observe at this point
that the fact that we can fit data within modified stochastic approach only by
allowing for a kind of "quasiparticles" of mass $m$, different for different
centralities, corresponds in a sense to introducing parameter $q$ to the usual
statistical model. The possible  dynamical origin and meaning of such variable mass
is, however, still lacking.}.

%
% SECTION-4
%
\section{Concluding remarks} \label{sec:4}

We have provided here systematic analysis of recent RHIC data on $k_t$ distributions
\cite{Arsene:2003yk,Adams:2003kv,Adcox:2002pe} by using three different kinds of
statistical approaches: Hagedorn model \cite{Hagedorn:1965aa}, two versions of the
modified statistical based on Tsallis statistics \cite{Tsallis:1988aa} and a suitable
adaptation of the stochastic model proposed in \cite{Biyajima:2002at}. We have found
that Hagedorn-type model
\begin{figure}
  \resizebox{0.50\textwidth}{!}{\includegraphics{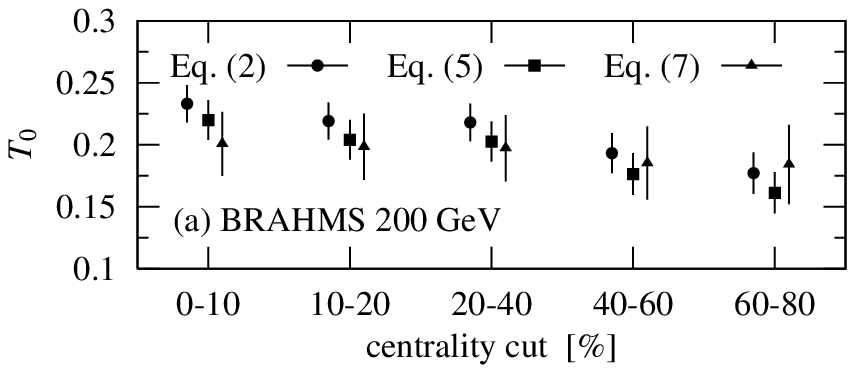}}
  \resizebox{0.50\textwidth}{!}{\includegraphics{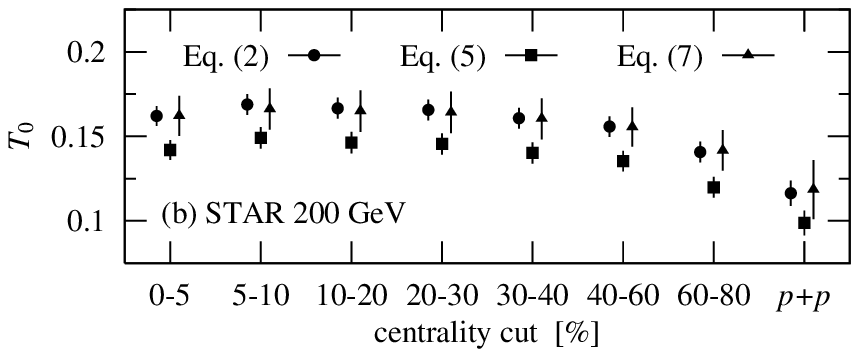}}
  \resizebox{0.50\textwidth}{!}{\includegraphics{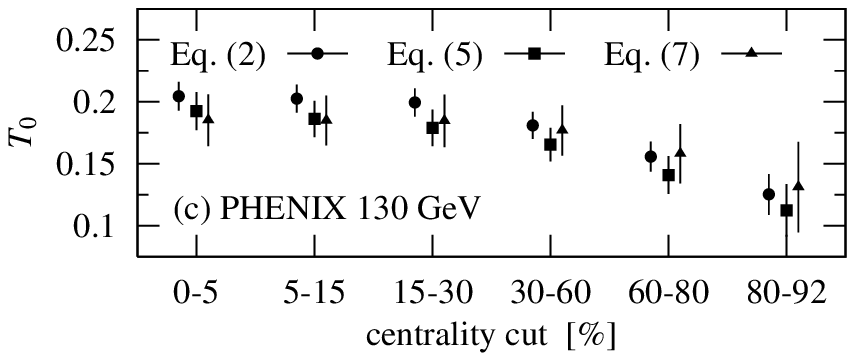}}
  \caption{Comparison of temperatures of hadronization obtained by
           using different approaches as given by: $(a)$ - Eq.
           \eqref{eq:2} with $Q=q/(1-q)$; $(b)$  - Eq. \eqref{eq:2}
           and $Q=1/(1-q)$; $(c)$ - Eq. \eqref{eq:4}. In the later
           case $T_0$ has been obtained from the values of $V_t^2$
           and $m$ obtained in Fig.~\ref{fig:3} by using Einstein's
           relation: $T_0=m\cdot V^2(t)$.}
  \label{fig:4}
\end{figure}
\cite{Hagedorn:1965aa} cannot fit data at large $k_t$ (its widely used for quick
estimations simplified version with $\rho(m) =1$, which is then just a simple
Boltzmann gas model with only one parameter, the temperature $T_0$, fails completely
even for smaller $k_t$, cf. Table~\ref{table:brahms}). However, these data can still
be reasonably well fitted either by non-extensive extensions of statistical model
\cite{Tsallis:1988aa} or by picture of some suitable diffusion process taking place
in transverse rapidity space \cite{Biyajima:2002at,B3}. This is specially true if one
limits itself to $k_t<5$ GeV/c range as the case of BRAHMS \cite{Arsene:2003yk} and
PHENIX \cite{Adcox:2002pe} data, the $k_t <12$ GeV/c range considered in STAR
experiment \cite{Adams:2003kv} seems to be already too big to be fitted properly even
with these two approaches (the corresponding values of the $\chi^2$ are considerably
bigger in this case and the values of parameters obtained for STAR and BRAHMS data,
which were taken at the same collision and at the same energy, are also different).

%%% TABLE 3 %%%

\begin{table*}
\caption{Comparison of investigated models: simple statistical model (i.e., Hagedorn
model as given by eq. (\ref{eq:1}) but with $\rho(m)=1$, in which case it is just a
simple statistical Boltzmann gas model with only one parameter, namely temperature
$T_0$), non-extensive Tsallis distribution (NETD) and Ornstein-Uhlenbeck process
(O-U), using data on $k_t$ distributions at $\sqrt{s_{NN}} = 200$ GeV obtained by
BRAHMS Collaboration \cite{Arsene:2003yk} for smallest and largest centralities.}
\begin{center}
\begin{tabular}{|cc|ccc|ccc|ccc|}
\hline
           & & \multicolumn{3}{|c|}{Simple statistical model, }&
\multicolumn{3}{|c|}{NETD Eq.~\eqref{eq:2}}
&\multicolumn{3}{|c|}{ O-U Eq.~\eqref{eq:4}}\\
           & & \multicolumn{3}{|c|}{Eq.~\eqref{eq:1} with $\rho(m)=1$} &
\multicolumn{3}{|c|}{(with $Q=q/(1-q)$)}&\multicolumn{3}{|c|}{}\\
%\hline
C.C (\%)   & & $T_0$ & $q$ & $m$   &  $T_0$ & $q$ & $m$   & $T_0$ &
$q$ & $m$ \\
           & & (GeV) &     & (GeV) &  (GeV) &     & (GeV) & (GeV) &
& (GeV)\\
\hline
0-10  & $\chi^2$/n.d.f& \multicolumn{3}{|c|}{177/23} &
\multicolumn{3}{|c|}{10.2/23} &       \multicolumn{3}{|c|}{39.9/23}\\

      &  & 0.302   & --- & --- & 0.232 & 1.043 & --- & 0.201 & --- & 0.784 \\
\hline
60-80 & $\chi^2$/n.d.f& \multicolumn{3}{|c|}{567/22} &
\multicolumn{3}{|c|}{2.76/22} &       \multicolumn{3}{|c|}{4.06/22}\\
      &  & 0.325   & --- & --- & 0.175 & 1.084 & --- & 0.184 & --- & 0.515\\
\hline
\end{tabular}
\label{table:brahms}
\end{center}
\end{table*}

As is shown in Fig. \ref{fig:4}, the temperatures $T_0$ obtained in modified
statistical and stochastic approaches (with varying mass $m$) are essentially very
similar to each other and follow the same dependence on the centrality, namely $T_0$
decreases when collision is more peripheral. However, because stochastic approach
seems to be more dynamical than $q$-statistical one (where the true dynamical origin
of the nonextensivity parameter is not yet firmly established, see
\cite{Wilk,Wilk:1999dr}), we regard as the most valuable our finding that stochastic
approach \cite{Biyajima:2002at,B3,SB} works so well and can serve to provide first
simple estimations of any new data in the future. On the other hand we have also
demonstrated that the two possible approaches using $q$-statistics are equivalent to
each other, at least in the frame of limited phenomenological approach presented
here. One should also stress at this point that $q$ statistical approach offers
unique information on fluctuations in the system, which can be translated into
information on its volume. Our results for $AA$ and $pp$ collisions taken together
with old results for $e^+e^-$ annihilations indicate in this respect distinct growth
of the expected volume of interactions from the most elementary annihilation
processes to the nuclear collisions.

%
% ACKNOWLEDGEMENT
%
\section*{Acknowledgements}
We are indebted to various conversations with participants at the RITP meeting (Kyoto
University, Aug. 2003) and at the RCNP meetings (Osaka University, Oct. 2003 and
November 2004). We also appreciate information on Ref. \cite{MC} provided us by W.
Zajc. This study is partially supported by Faculty of Science at Shinshu University.
Partial support of the Polish State Committee for Scientific Research (KBN) (grants
621/E-78/SPUB/CERN/P-03/DZ4/99 and 3P03B05724 (GW)) is also acknowledged.

\end{document}